\documentclass{article}
\usepackage{icomp2024_conference,times}
\usepackage{algpseudocode}
\usepackage{graphicx}
\usepackage{amsmath}
\usepackage{siunitx}
\usepackage{subcaption} 

\usepackage{amsmath,amsfonts,bm}

\def\eqref#1{equation~\ref{#1}}

\def\1{\bm{1}}

\DeclareMathAlphabet{\mathsfit}{\encodingdefault}{\sfdefault}{m}{sl}
\SetMathAlphabet{\mathsfit}{bold}{\encodingdefault}{\sfdefault}{bx}{n}

\newcommand{\R}{\mathbb{R}}

\usepackage{hyperref}
\usepackage{amsmath, accents}
\usepackage{url}
\usepackage{algorithm}

\title{Exploring Applications of State Space Models  and Advanced Training Techniques in Sequential Recommendations: A Comparative Study on Efficiency and Performance}

\author{Makar Baderko\thanks{These authors contributed equally to this work.} \\
\texttt{makarbaderko@gmail.com} \\
\And Stepan Kulibaba\footnotemark[1] \\ \texttt{kulibabast@gmail.com} \\
\And
Mark Obozov\footnotemark[1] \\
\texttt{obozovmark9@gmail.com} \\
\And
Nikolay Kutuzov \\
\texttt{kutuzov.nv@phystech.edu}
\And Alexander Gasnikov \\
\texttt{gasnikov@yandex.ru} \\
}

\icompfinalcopy 
\begin{document}
\maketitle
\begin{abstract}
Recommender systems aim to estimate the dynamically changing user preferences and sequential dependencies between historical user behaviour and metadata. Although transformer-based models have proven to be effective in sequential recommendations, their state growth is proportional to the length of the sequence that is being processed, which makes them expensive in terms of memory and inference costs. Our research focused on three promising directions in sequential recommendations: enhancing speed through the use of State Space Models (SSM), as they can achieve SOTA results in the sequential recommendations domain with lower latency, memory, and inference costs, as proposed by \cite{mamba4rec}; improving the quality of recommendations with Large Language Models (LLMs) via Monolithic Preference Optimization without Reference Model (ORPO); and implementing adaptive batch- and step-size algorithms to reduce costs and accelerate training processes.
\end{abstract}

\section{Introduction}

Modern digital products are heavily dependent on the performance of the recommender systems, that are frequently utilized to predict future interactions of customers based on their historical behaviour.

While recurrent neural networks (RNNs) and convolutional neural networks (CNNs) led the way in the use of neural networks in sequential recommendation, they suffer from the forgetting issue \cite{Kirkpatrick_2017}. Recently, researches were able to introduce the Transformer-based models into the field \cite{GPT4Rec}.

Due to the quadratic computational cost, these attention-based approaches typically suffer from the inference inefficiency problem despite their impressive performance.

SSM models have been adopted as a replacement for RNNs and CNNs, with \cite{mamba4rec} being an example. We develop this idea further, while introducing a novel state-of-the-art sequential recommender model Hydra4Rec, based on the architecture proposed in \cite{hydra}.

\subsection{Related study}

\subsubsection{Sequential recommendation task definition}

Consider the user set $U = \{u_1, u_2, ..., u_N\}$, item set $V = \{v_1, v_2, ..., v_K\}$ and $S_u = \{v_1, v_2, ..., v_{n_u}\}$ as the chronologically ordered interaction sequence for user $u \in U$, where $n_u$ is the the length of the sequence. Given $S_u$ the task is to predict the next
interacted item, $v_{n_u + 1}$.

\subsubsection{Transformers}

Recently, transformer models have been shown to be effective in sequential recommendation tasks as the backbone of larger models \cite{SASRec} and as individual LLMs \cite{GPT4Rec}, \cite{LlamaRec}. Despite their success, attention-based methods face inference inefficiencies due to the quadratic computational complexity inherent in attention operators and their rapid state growth, which is proportional to the sequence length. Also, without special mechanisms \cite{RingAttention}, transformers can't handle long contexts and, consequently, long user histories.

\subsubsection{State Space Models}

State Space Model (SSM) is a recent framework for sequence modelling defined by linear ordinary differential equations:

\begin{align*}
h'(t) = Ah(t) + Bx(t) \\
y(t) = Ch(t) 
\end{align*}

Where $A,B,C$ are learnable matrices, $h(t)$ is the latent space, $x(t)$ is the input sequence and $y(t)$ is the output sequence. To compute sequence-to-sequence transformations efficiently, the matrix $A$ must be \textit{structured}, so structured SSMs have been introduced \cite{StructuredGu}. A general form of a structured SSM is defined by the equations:

\begin{align*}
    h_t = Ah_{t-1} + Bx_t \\
    y_t = Ch_t
\end{align*}

Where $A \in \R^{(N,N)}, B \in \R^{(N,1)}, C \in \R^{(N,1)}$. They map a 1-dimensional sequence $x \in \R^T \rightarrow y \in \R^T$ through an implicit latent state $h \in \R^{(T,N)}$. To operate directly on sequences, the discretisation rule  $(f_{A}, f_{B})$  is applied to continuous parameters $(\Delta, \mathring{A}, \mathring{B})$, by $A = f_{A}(\Delta, \mathring{A})$ and $B = f_{B}(\Delta, \mathring{B})$, where $\Delta$ is the parameterised step size.

\subsubsection{Mamba block}

In order to adaptively focus on relevant information while filtering out noise, the Mamba block \cite{mamba} introduces an extension to structured SSMs by adding a data-dependent selection mechanism. An important feature of Selective SSMs is their ability to be computed efficiently on the GPU using kernel fusion, parallel scanning and recomputation mechanisms. 

In contrast to transformers' $O(n^2 \cdot d)$, SSMs provide $O(n \cdot d^2)$ complexity, which makes them a more efficient alternative, especially when operating on long sequences (\cite{mamba2}) .

\subsubsection{Mamba applications to sequential recommendations}

Several applications of Mambas selective SSMs to recommender systems have been introduced. \cite{mamba4rec}, \cite{echomamba4rec}. As Mamba has already shown efficient results in different areas \cite{visionmamba} \cite{audiomamba} \cite{timemamba}, we propose a more complex exploration of Mamba applications to recommender systems.

\subsection{Hydra}

Hydra fully utilize the matrix mixer framework, to explore a novel bidirectional sequence mixer. To achieve linear complexity, special class of quasiseparable matrices is considered. A matrix $\textbf{M}$ is quasiseparable if every label element $m_{ij}$ satisfies:

\begin{equation}
        \label{eq:quasi}
        m_{ij} =
        \begin{cases}
            \overrightarrow{\textbf{c}^{T}_{i}} \overrightarrow{\textbf{A}^{\times}_{i:j}} \overrightarrow{\textbf{b}_{j}},  & \text{if } i > j \\
            \delta_{i},         & \text{if } i = j \\
            \overleftarrow{\textbf{c}^{T}_{j}} \overleftarrow{\textbf{A}^{\times}_{j:i}} \overleftarrow{\textbf{b}_{i}},  & \text{if } i < j\\
        \end{cases},
\end{equation}

\noindent
where each $\delta_i$ is a scalar, $\mathbf{b}_i, \mathbf{c}_i \in \mathbb{R}^{N \times 1}$, and $\mathbf{A}_i \in \mathbb{R}^{N \times N}$.
\noindent
Quasiseparable matrices can be used as matrix mixers, this framework is named as \textbf{Hydra}, achieving $O(L)$ complexity. Based on an important property of quasiseparable matrices - quasiseparable matrices can be expressed as a combination of two semiseparable matrices - a subquadratic matrix multiplication algorithm was developed. Then, as a semiseparable matrix structure includes SSMs \cite{mamba2}, any SSM could be combined with the Hydra framework, resulting in a new efficient bidirectional SSM architecture.

\subsubsection{Universal Gradient Method for Stochastic Optimization}

The release of the Universal Stochastic Gradient Method (USGM) by \cite{rodomanov2024universalgradientmethodsstochastic} represents a noteworthy development in the field of stochastic optimization, that is the backbone of machine learning.

The classical Stochastic Gradient Descent (SGD) by \cite{Robbins1951ASA} is one of the earliest techniques in the field of stochastic optimization. Its novelty at the time of publication helped it gain popularity. However, because hyperparameter values are only defined prior to the start of the training phase, SGD achieves poor convergence rates in real-world applications. 

A number of SGD-based adaptive strategies, including Adam by \cite{kingma2017adammethodstochasticoptimization} and Adagrad by \cite{adagrad}, have been put forth to address the issue. Although they are more efficient than SGD, because they modify the step size according to the aggregate of the previous gradients, when applied to non-convex optimization problems, they might still experience difficulty achieving convergence.

The USGM develops the idea of adaptiveness further, as the algorithm can adapt to the curvature of the loss function, thus providing more accurate gradient estimates even in highly noisy environments. Classic optimization algorithms are typically designed to handle either smooth or non-smooth problems. USGM, however, is an algorithm capable of addressing both classes of problems and adapting to oracle noise with unknown variance. The algorithm can adjust to Hölder-Lipschitz coefficients without requiring precise calculation of those. The only hyperparameters used by the algorithm are the upper bound on the diameter of the search hypersphere, the initial learning rate and batch size, what makes it a more practical choice for real-world applications.

$\mathbf{Notation}$

The same mathematical notation as in the original USGM paper will be used.

The loss function is defined as $f$ and the parameters of our model are defined as $x$.

To characterize the smoothness of $f$ the H\"older constant is introduced for each $\nu \in [0, 1]$:
\begin{equation}
L_{\nu} := \sup_{\substack{x,y \in \text{Dom}f, x \neq y, \\ g(x) \in \partial f(x), g(y) \in \partial f(y)}} \frac{\|g(x) - g(y)\|_*}{\|x - y\|^{\nu}} , \tag{1.1}
\end{equation}

where $g(x)$ and $g(y)$ are the stochastic approximations of subgradients of our loss function $f$ at points $x$ and $y$ respectively.

The dual norm is defined in the standard way: 

\begin{equation}
\|s\|_* := \max_{\|x\|=1} \langle s, x \rangle = \langle s, B^{-1} s \rangle^{1/2}, \quad s \in \mathbb{R}^n. \tag{1.2}
\end{equation}

Of course, for certain values of the exponent $\nu \in [0, 1]$, it may happen that $L_{\nu} = +\infty$. However, we assume that there exists (at least one) exponent for which the corresponding H\"older constant is finite.

For any $ t \in \mathbb{R} $, by $ [t]_+ := \max\{t, 0\} $, its positive part is denoted. For random variables $ X $ and $ \xi $, by $ \mathbb{E}_\xi[X] $ and $ \mathbb{E}[X] $, the expectation of $ X $ w.r.t. $ \xi $, and the full expectation of $ X $, respectively are denoted.

\begin{algorithm}[H]
\caption{Universal Stochastic Gradient Method}
\begin{algorithmic}[1]
\State \textbf{Initialize:} $x_0 \in \text{dom} \, f$, $D > 0$, $H_0 := 0$, $g_0 \sim \hat{g}(x_0)$.
\For{$k = 0, 1, \dots$}
    \State $x_{k+1} = \arg\min_{x \in \text{dom} \, f} \{ \langle g_k, x \rangle + \frac{H_k}{2} \| x - x_k \|^2 \}$.
    \State $g_{k+1} \sim \hat{g}(x_{k+1})$.
    \State $H_{k+1} := H_k + \frac{\left[\hat{\beta}_{k+1} - \frac{1}{2}H_kr_{k+1}^2\right]_{+}}{D^2 + \frac{1}{2}r_{k+1}^2}$,
    \State where $r_{k+1} = \| x_{k+1} - x_k \|$, $\hat{\beta}_{k+1} = \langle g_{k+1} - g_k, x_{k+1} - x_k \rangle$.
\EndFor
\end{algorithmic}
\end{algorithm}

The authors of the paper also define the upper bound for the stochastic approximation of the
symmetrized Bregman distance for points $x_k$ and $x_{k+1}$ (${\hat{\beta}}_{k+1}$) as follows:

\begin{equation}
\hat{\beta}_{k+1} = \left\langle f'(x_{k+1}) - f'(x_k) + \Delta_{k+1}, x_{k+1} - x_k \right\rangle \leq L_{\nu} r_{k+1}^{1+\nu} + \sigma_{k+1} r_{k+1}, \tag{1.3} \label{eq:1.3}
\end{equation}

where $f'(x_k) := \mathbb{E}_{\xi_k}[g_k] \in \partial f(x_k)$, $\Delta_{k+1} := \delta_{k+1} - \delta_k$ with $\delta_k := g_k - f'(x_k)$ being the error of the stochastic gradient (such that $\mathbb{E}\|\delta_k\|^2 \leq \sigma^2$), and $\sigma_{k+1} := \|\Delta_{k+1}\|$. 

\section{Odds Ratio Preference Optimization}
Odds Ratio Preference Optimization (\texttt{ORPO}), \cite{ORPO} - is novel preference optimization framework that consolidates an odds ratio-based penalty. 

The odds of generating the output sequence $y$ given an input sequence $x$ are defined by:

\begin{equation}
    \log P_\theta(y|x) = \frac{1}{m} \sum_{t=1}^m \log P_\theta(y_t|x, y<t)
\end{equation}

\begin{equation}
    \textbf{odds}_\theta(y|x) = \frac{P_\theta(y|x)}{1 - P_\theta(y|x)}
\end{equation}

The odds ratio of the chosen response $y_w$ over the rejected response $y_l$, $\textbf{OR}_\theta(y_w, y_l)$, indicates how much more likely it is for the model $\theta$ to generate $y_w$ than $y_l$ given input $x$.

\begin{equation}
    \textbf{OR}_\theta(y_w, y_l) = \frac{\textbf{odds}_\theta(y_w|x)}{\textbf{odds}_\theta(y_l|x)}\label{eq:or}
\end{equation}

\subsection{Objective Function of \texttt{ORPO}}

The objective function of \texttt{ORPO} consists of two components: 1) supervised fine-tuning (SFT) loss ($\mathcal{L}_{SFT}$); 2) relative ratio loss ($\mathcal{L}_{OR}$).
\begin{equation}
    \mathcal{L}_{ORPO} = \mathbb{E}_{(x, y_w, y_l)}\left[ \mathcal{L}_{SFT} + \lambda \cdot \mathcal{L}_{OR} \right]
\end{equation}

$\mathcal{L}_{SFT}$ follows the conventional causal language modelling negative log-likelihood (NLL) loss function to maximise the probability of generating the reference tokens.
The $\mathcal{L}_{OR}$ in the equation maximises the odds ratio between the likelihood of generating the unfavourable response $y_w$ and the favourable response $y_l$. 

\begin{equation}
    \mathcal{L}_{OR} = -\log \sigma \left( \log \frac{\textbf{odds}_\theta(y_w|x)}{\textbf{odds}_\theta(y_l|x)} \right) \label{eq:ratio} 
\end{equation}
Together, $\mathcal{L}_{SFT}$ and $\mathcal{L}_{OR}$ weighted with $\lambda$ tailor the pre-trained language model to adapt to the specific subset of the desired domain and disfavor generations in the rejected response sets.

\subsection{Gradient of \texttt{ORPO}}

The gradient of $\mathcal{L}_{Ratio}$ further justifies the use of the odds ratio loss. It consists of two terms: one that penalises wrong predictions and one that contrasts between selected and rejected answers, for $d=(x, y_l, y_w)\sim D$.

\begin{equation}
    \nabla_\theta \mathcal{L}_{OR} = \delta(d) \cdot h(d)\label{eq:simp}
\end{equation}

\begin{align}
    \delta(d) &= \left[ 1 + \frac{\textbf{odds}_\theta P(y_w|x)}{\textbf{odds}_\theta P(y_l|x)} \right]^{-1}\label{eq:delta} \\
    h(d) &= \frac{\nabla_\theta \log P_\theta(y_w|x)}{1 - P_\theta(y_w|x)} - \frac{\nabla_\theta \log P_\theta(y_l|x)}{1 - P_\theta(y_l|x)}\label{eq:grad}
\end{align}

If the probabilities of the favoured responses are relatively higher than the disfavoured responses, $\delta(d)$ in the equation will converge to 0. This indicates that the $\delta(d)$ will play the role of a penalty term, speeding up the parameter updates when the model is more likely to generate the disfavoured responses. 

Meanwhile, $h(d)$ implies a weighted contrast of the two gradients from the chosen and rejected responses. Specifically, $1-P(y|x)$ in the denominators amplifies the gradients when the corresponding side of the likelihood $P(y|x)$ is low. As the likelihood increases, the model adapts more quickly to the distribution of selected responses.

\section{Our results}

\subsection{2Mamba4Rec}

To compare the Mamba2 model more fairly with its predecessor (Mamba1) we used the model architecture from the Mamba4Rec article by \cite{mamba4rec}, replacing the Mamba block with the Mamba2 block introduced by \cite{mamba2}.

\subsection{MamRec}

We implemented the standard GPT4Rec architecture from \cite{GPT4Rec} while replacing the GPT-2 model \cite{gpt2} with a Mamba2 \cite{mamba2} model in order to achieve better performance. 

\subsection{Hydra layer}

To apply the sequence-to-sequence potential of the Hydra block \cite{hydra}, we provide a custom Hydra layer that combines the Hydra block with a standard feed-forward network. The main part of our standard architecture consists of Hydra layers. We have found that Hydra layers work more effectively than standard Mamba layers in several cases. Then we use the same PFFN that was introduced in Mamba4Rec \cite{mamba4rec} \ 

\[
PFFN(H) = GELU(HW^{(1)} + h^{(1)})W^{(2)} + b^{(2)}
\]

\noindent
Where $W^{(1)} \in \mathbb{R}^{D \times 4D}$, $W^{(2)} \in \mathbb{R}^{D \times 4D}$,
$b^{(1)} \in \mathbb{R}^{D}$ are parameters of two dense layers and we use GELU activation.

\subsubsection{Prediction layer}

Prediction layer is adopted from SASRec and Mamba4Rec, last item embedding is used to generate the final prediction scores:

\[
\hat{y} = Softmax(hE^T) \in \mathbb{R}^{|V|}
\]
\noindent
where $h \in \mathbb{R}^D$ is the last item embedding from the Hydra layer and $\hat{y} \in \mathbb{R}^{|V|}$ represents the probability distribution over the next item
in the item set $V$.

\subsection{Benchmarks}
We tested our models on 3 benchmarks: Amazon Reviews '23 Beauty and Personal care, Amazon Reviews '23 Video Games and MovieLens-1M. 

All experiments were conducted using the same initial hyperparameters for better interpretability. 

The number listed in parentheses after the number of parameters of each models is the number of parameters after deducting the size of all embedding layers if present.

\subsubsection{Latency}


For the LLMs, the measure tokens-per-second was selected, whilst the average time to generate suggestions for a single user was chosen to compare the others.

Since all of the latency values that we are dealing with are relatively small, it is important to ascertain the measurement's margin of error in order to comprehend the experimental results that were produced.

Consequently, the bootstrap sampling method was applied. 30 samples comprising 1500 users randomly chosen from the original dataset were used to individually measure the latency values. Following the removal of outliers, the distribution of the latencies was examined, and the confidence intervals were computed.



The latency measurements were performed on one Nvidia A100-80GB graphics card.

\subsubsection{Interpretation of results}
It is evident that Hydra shows metrics that are similar, sometimes better to those of Mamba, but its latency is four to five times lower. Simultaneously, models such as LlamaRec exhibit superior performance; however, their substantial parameter count may constrain their applicability in real-world situations. Metrics-wise and latency-wise, SSM-based models outperform the baseline SASRec model on average.

According to the study, SSMs can outperform LLMs while still operating at a greater speed, since they require less number of parameters.

\begin{table}[htbp]
\caption{Amazon Reviews '23 Beauty and Personal care}
\label{beauty-table}
\begin{center}
\begin{tabular}{lccccccc}
\multicolumn{1}{c}{\bf Model}  &\multicolumn{1}{c}{\bf HT@10} &\multicolumn{1}{c}{\bf NDCG@10} &\multicolumn{1}{c}{\bf MRR@10}
&\multicolumn{1}{c}{\bf Latency} &\multicolumn{1}{c}{\bf \# Parameters}  
\\ \hline \\


SASRec &{0.048} &{0.028} &{0.022} &{0.117} &{14M \ (100k)} \\
Mamba4Rec &{0.048} &{0.030} &{0.025}  &{0.0075} &{14M \ (80k) }\\
MamRec &{0.031} &{0.020} &{0.017}  &{2.51} &{130M} \\
GPT4Rec &{0.030} &{0.025} &{0.015}  &{2.32} &{117M} \\
2Mamba4Rec &{0.048} &{0.031} &{0.028}  &{0.0110} &{14M \ (80k) }\\
LlamaRec &{\textbf{0.093}} &{0.040} &{\textbf{0.040}}  &{-} &{7B} \\
Hydra4Rec &{0.070} &{\textbf{0.043}} &{{0.035}}  &{\textbf{0.0042}} &{840k} \\

\end{tabular}
\end{center}
\end{table}

\begin{table}[htbp]
\caption{Amazon Reviews '23 Video Games}
\label{games-table}
\begin{center}
\begin{tabular}{lccccccc}
\multicolumn{1}{c}{\bf Model}  &\multicolumn{1}{c}{\bf HT@10} &\multicolumn{1}{c}{\bf NDCG@10}  &\multicolumn{1}{c}{\bf MRR@10} &\multicolumn{1}{c}{\bf Latency} &\multicolumn{1}{c}{\bf \# Parameters} 
\\ \hline \\


SASRec &{0.119} &{0.073} &{0.059}  &{0.129} & 1.8M \ (100k) \\
Mamba4Rec &{0.107} &{0.062} &{0.048}  &{0.0088} &{1.8M \ (80k)}\\
MamRec &{0.083} &{0.033} &{0.025}  &{2.51} &{130M} \\
GPT4Rec &{0.080} &{0.042} &{0.026}  &{2.32} &{117M} \\
2Mamba4Rec &{0.118} &{0.061} &{0.048}  &{0.0103} &{1.8M \ (80k)}\\
LlamaRec &{\textbf{0.150}} &{\textbf{0.098}} &{\textbf{0.064}}  &{-} &{7B} \\
Hydra4Rec &{0.112} &{0.059} &{0.044}  &{\textbf{0.0037}} &{750k} \\

\end{tabular}
\end{center}
\end{table}

\begin{table}[htbp]
\caption{MovieLens-1M}
\label{ml-table}
\begin{center}
\begin{tabular}{lccccccc}
\multicolumn{1}{c}{\bf Model}  &\multicolumn{1}{c}{\bf HT@10} &\multicolumn{1}{c}{\bf NDCG@10} &\multicolumn{1}{c}{\bf MRR@10} &\multicolumn{1}{c}{\bf Latency} &\multicolumn{1}{c}{\bf \# Parameters}
\\ \hline \\

SASRec &{0.224} &{0.117} &{0.084}  &{0.128} &{320k \ (100k)}\\
Mamba4Rec &{0.303} &{0.178} &{0.139}  &{0.0020} &{300k \ (80k)} \\
MamRec & 0.201 & 0.072 & 0.064  &{2.51} &{130M} \\
GPT4Rec &{0.212} &{0.074} &{0.060}  &{3.23} &{117M} \\
2Mamba4Rec &\textbf{{0.340}} &\textbf{{0.193}} &\textbf{{0.148}} &{0.0034} &{300k \ (80k)} \\
LlamaRec &{0.148} &{0.067}  &{-} &- &7B \\
Hydra4Rec &{0.308} &{0.179} &{0.140} & {\textbf{0.0005}} &{290k}  \\

\end{tabular}
\end{center}
\end{table}

\newpage

\subsection{Application of ORPO to recommender systems}

To improve quality and incorporate human preference into the LLM, we used the ORPO technique described above. Main motivation is ability to slightly improve quality of the large model, showing the potential of preference optimization techniques in recommeder systems.

\subsubsection{Baseline}

As a baseline, we considered the standard LlamaRec pipeline with Llama 7B LLM. Both retrieval and LLM were fine-tuned as described in the LlamaRec paper prior to the ORPO procedure.

\subsubsection{Data}

We used the standard LlamaRec format. To construct the text input, we prepend an instruction to describe the task, followed by both history and candidate
items, represented by their titles. Our prompt template is:
"
\textit {Instruction}: Given the user's history in chronological
order, recommend an item from the candidate pool
with its index letter;
\textit {Input}: User history: { history }; \textit {Candidate pool}:
{ candidates };
\textit{Answer}: { label }
"
where history, candidates and label are replaced by history
item titles, candidate item titles and the labeled of each data item. For inference, the label position is left empty for the model to provide predictions.

We then created a pair data set, where each pair consists of a winner and a loser. It consists of 2 parts. The first part was created from the last 2 items from the user history, where the winner is the item with the higher rating and the loser the opposite. To create the second part, we used negative sampling training LightFM model and took the last user's movie as the winner and the worst recommendation of LightFM as the loser.

\subsubsection{Experimental setup}

For the ORPO procedure we used lr = 8e-6, ORPO uses very low learning rates compared to traditional SFT or even DPO. This value of 8e-6 comes from the original paper. beta = 0.1  An appendix from the original paper shows how it's been selected with an ablation study. We use paged adamw optimizer with gradient accumulation steps = 4. We train for 1 training epoch and evaluate each 0.2 steps, also we set warmup steps to 10. We use linear scheduler type. Other parameters are dependant on capabilities of your resources. Training was proceed on 1 A100 80GB videocard.


\subsubsection{ORPO results}

\begin{figure}[h!]
    \centering
    \begin{minipage}[b]{0.5\linewidth}
        \centering
        \includegraphics[width=0.6\linewidth]{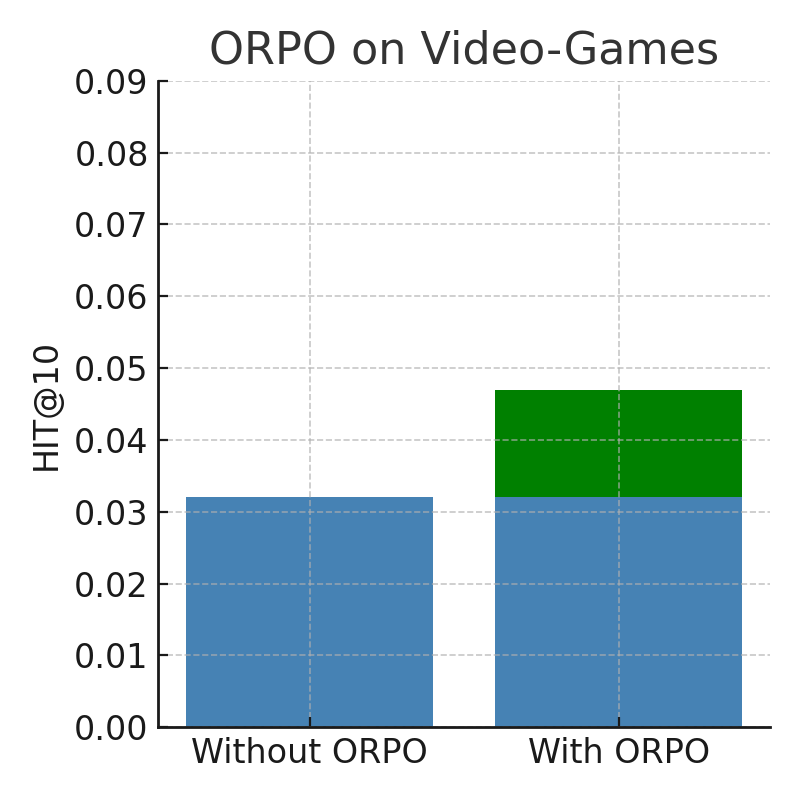}
        \caption{Video-Games}
        \label{fig:video-games}
    \end{minipage}%
    \begin{minipage}[b]{0.5\linewidth}
        \centering
        \includegraphics[width=0.6\linewidth]{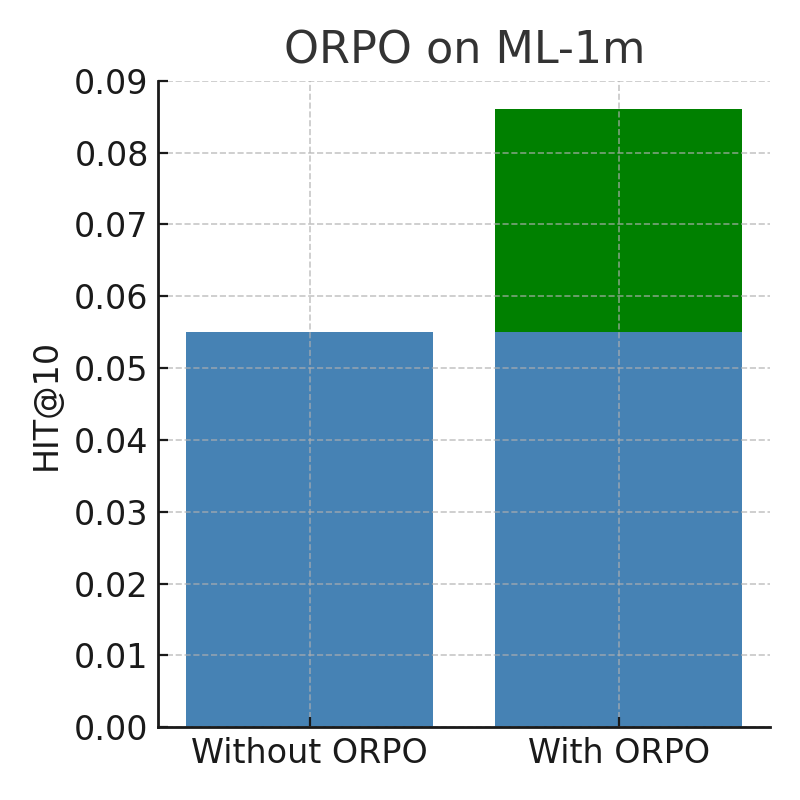}
        \caption{ML-1m}
        \label{fig:ml1m}
    \end{minipage}
\end{figure}

ORPO procedure provide slight impact on LLM perfomace not affecting original latency. See Figure 2 and 3.

\subsection{Adaptive Batching}

\subsubsection{Algorithm}

We were able to apply the concept of adaptive batching to the aforementioned algorithm (USGM) and enhance its performance in training SSMs in the sequential recommendations domain. The approach presented in this section is novel, since adaptive batching techniques have never been applied to the \cite{rodomanov2024universalgradientmethodsstochastic}.

In stochastic optimization, the variance of the gradient estimates reduces as the batch size increases. Specifically, the variance decreases proportionally to $\frac{1}{B}$. Concurrently, the standard deviation ($\sigma$) of the gradient estimates decreases at the rate of $\frac{1}{\sqrt{B}}$. From here \ref{eq:1.3} , we can derive the equation \ref{eq:2.1} Thus, the variance and standard deviation of the gradient estimations decrease as the batch size increases. But as the batch size increases, the impact of this reduction decreases. Additional bigger increases in batch size result in insignificant decreases in variance, as accuracy gains have landed on a plateau. Therefore, increasing the batch size further at this stage is not effective, since it will not provide any improvements in accuracy, but will require additional computational resources. 

\begin{equation}
\hat{\beta}_{k+1} \leq  L_{\nu} r_{k+1}^{1+\nu} + \frac{\sigma_{k+1} r_{k+1}}{\sqrt{B}}, \tag{2.1} \label{eq:2.1}
\end{equation}

Further into the paper we will consider this upper estimate an equality to indirectly calculate the values of $L_{\nu} r_{k+1}^{1+\nu}$ and $\sigma_{k+1} r_{k+1}$, since we know the exact value of $\hat{\beta}_{k+1}$ on each iteration of the algorithm.

As the coefficient $\hat{\beta}_{k+1}$ is computed on every iteration of the algorithm and the batch size is known, we can use the Weighted Least Squares method (with loss defined by the equation \ref{eq:2.2}) in linearized axes ($\frac{1}{\sqrt{B}}$ as the X axis and the $\hat{\beta}_{k+1}$ as the Y axis) to compute the values of $L_{\nu} r_{k+1}^{1+\nu}$ and $\sigma_{k+1} r_{k+1}$, the intercept and the slope of the linearized function respectively.

\begin{equation}
L = \sum_{k=0}^{K}(F(c_1, c_2, B_k) - \hat{\beta}_{k+1})^2 \cdot (1 - \alpha)^{K-k} \tag{2.2} \label{eq:2.2},
\end{equation}

where $F$ is the sought function $F = c_1 + \frac{c_2}{\sqrt{B}} \approx \hat{\beta}_{k+1} $, $c_1$ and $c_2$ are the values of $L_{\nu} r_{k+1}^{1+\nu}$ and $\sigma_{k+1} r_{k+1}$ respectively, predicted by the WLS algorithm. The value of alpha was empirically set to $\alpha = 0.01$, s.t.  the first points almost diminish. 

In the proposed algorithm, we will increase the batch size on each iteration ($B_k := B_{k-1} + B_0$) if $c_1 \neq \frac{c_2}{sqrt(B)}$ and recalculate the values of $c_1$ and $c_2$ using WLS ($L \underset{\substack{c_1, c_2}}{\rightarrow} \min$).

When we reach the plateau ($c_1 = \frac{c_2}{\sqrt{B}}$), we fix the current batch size until the end of the epoch. We consider this batch size perfect w.r.t the current conditions and label it as $B_i^{\ast}$.


When the epoch ends, we lower the batch size ($B_i := \frac{B_{i-1_{k}}}{\lambda}$), since the perfect batch size ($B_i^{\ast}$) could have changed since the last epoch. However, if it's bigger than $\frac{B_{i-1_{k}}}{\lambda}$, the algorithm will increase it during the next iterations. By doing so, we give the algorithm the ability to not continue training with excessive resource if the calculated batch size turns out to be too big in the current circumstances or to increase it even more if needed.

Since $B_k := B_{k-1} + B_0$, \ $B_k = m_k B_0  : \forall k \ (m_k \in \mathbb{Z})$. That means we can use the original Dataloader class from PyTorch (\cite{paszke2019pytorchimperativestylehighperformance}) wihout any modifications and take $m_k$ batches of initial size ($B_0$) on each iteration.

Experiments have shown that the use of values of beta and the batch size only from the current epoch in WLS almost do not change the desired Batch size ($B_i$) in comparison with using all of the previous values recorded during the training procedure. This proves that the values of $c1$ and $c2$ are, indeed, true for all parts of the loss landscape as we can see during the training procedure. This fact goes along well with the proposed theoretical assumptions, as these coefficients in part consist of values, that are constant for the whole function (the Hölder-Lipschitz constant $L_{\nu}$) and therefore should not change when we adjust the values of model's parameters $x$ during optimization.



\section{Discussion}

\textbf{Limitations.} ORPO and LLM related procedures require a lot of computational resources, especially for larger models. LLM inference itself also requires optimization, but this does not make them unusable in real tasks. SSM based models might have poor CPU performance due to their structure and implementation. Furthermore, optimizer that we proposed has poor robustness among its hyperparameters.
\newline
\noindent 
\textbf{Potential impact.} We believe that ideas and obtained results from our work can inspire the community
to continue the research in the field of recommender systems.
\newline
\noindent 
\textbf{Broader impact.} The goal of this paper is to advance the field of Machine Learning. There are many potential societal consequences of our work, none of which we feel must be specifically highlighted here.




\subsection*{Acknowledgments}
The paper was written from July 1 to July 24 as a part of the Sirius "Big Challenges" project programme in the "Big Data, Artificial Intelligence, Financial Technologies, and Machine Learning" category. \
We express our appreciation to MIPT, the Moscow Institute of Physics and Technology, for providing the computational resources required to finish this study.

\bibliography{icomp2024_conference}
\bibliographystyle{icomp2024_conference}

\appendix
\section{Appendix A. Datasets}

A table regarding the information about the datasets using in the paper is provided for scale. \ref{table:4}

\begin{table}[h]
\caption{Datasets}
\label{table:4}
\begin{center}
\begin{tabular}{lccccccc}
\multicolumn{1}{c}{\bf Datasets}  &\multicolumn{1}{c}{\bf Users} &\multicolumn{1}{c}{\bf Items} &\multicolumn{1}{c}{\bf Reviews}
\\ \hline \\
    Beauty and P.C.  & {750,835} & {211,452} & {6,860,059} \\
    MovieLens-1M  & {6,041} &  {3,417} & {999,611} \\
    Video Games  & {98,907} & {26,355} & {857,505} &  \\
\end{tabular}
\end{center}
\end{table}

\section{Appendix B. Benchmarks plots}

For a better understanding of our results, we propose plots with latency and the HIT@10 metric. Note that for better visualization, a logarithmic scale for latency is used.
\newpage
\begin{center}
\begin{figure}[t]
\begin{subfigure}{1\linewidth}
    \includegraphics[width=1\linewidth]{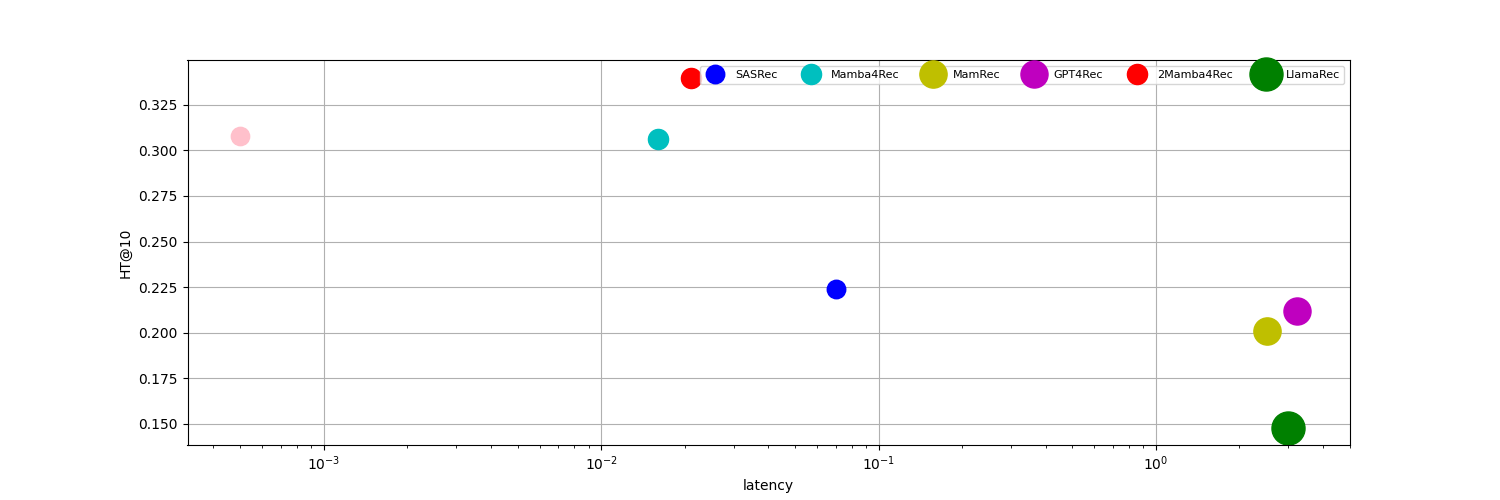}
    \caption{Comparison of performance of different models on the Beauty dataset.}
    \label{fig:beauty-label}
\end{subfigure}
\begin{subfigure}{1\linewidth}
    \includegraphics[width=1\linewidth]{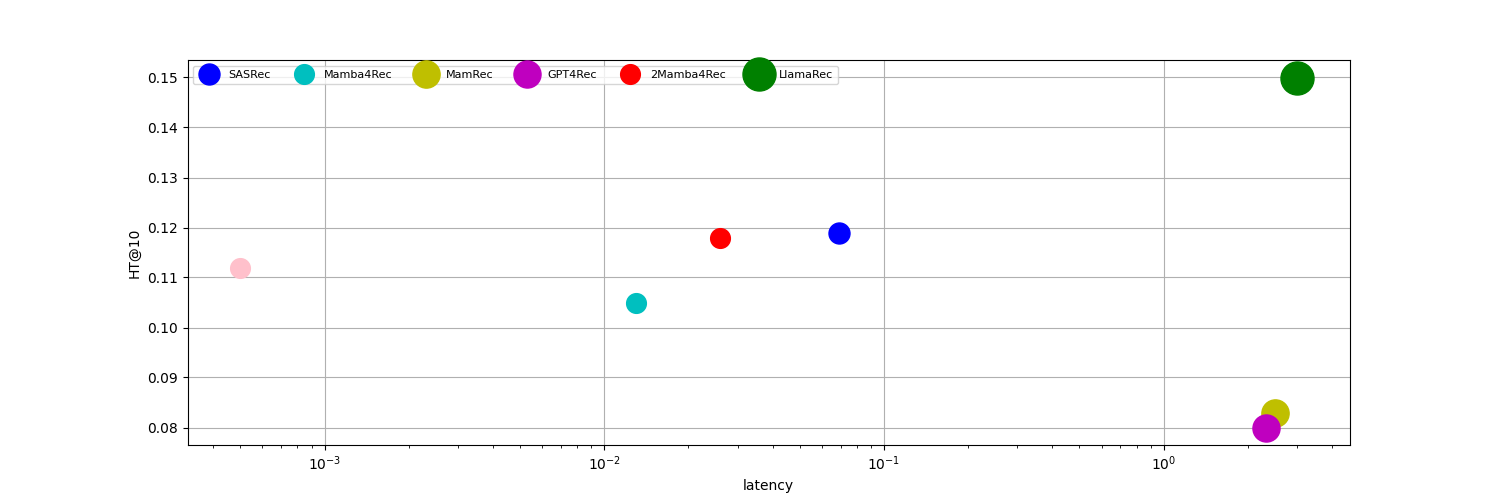}
    \caption{Comparison of performance of different models on the MovieLens-1M dataset.}
    \label{fig:ml-label}
\end{subfigure}
\begin{subfigure}{1\linewidth}
    \centering
    \includegraphics[width=1\linewidth]{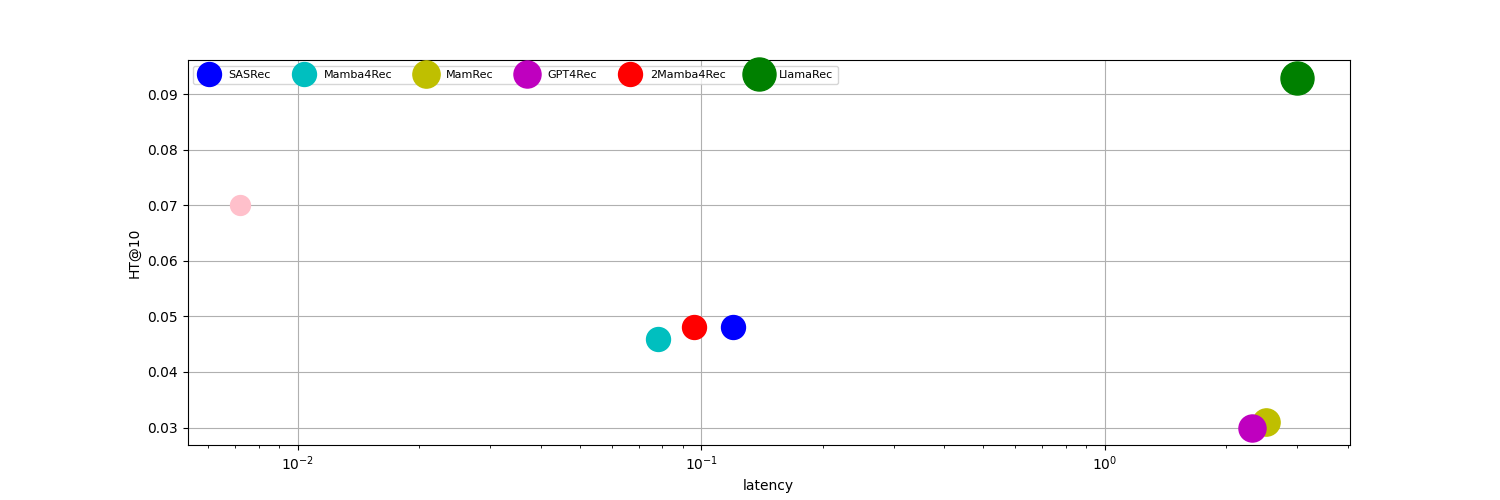}
    \caption{
Comparison of performance of different models on the VideoGames dataset.
}
    \label{fig:games-label}
\end{subfigure}
\end{figure}
\end{center}

\end{document}